\documentclass[12pt]{article}%
\usepackage{amsmath}
\usepackage{graphicx}
\usepackage{amsfonts}
\usepackage{amssymb}%
\providecommand{\U}[1]{\protect\rule{.1in}{.1in}}
\begin{document}

\title{Thermal expansion of the earth and the speed of neutrinos}
\author{C. S. Unnikrishnan\\\textit{Gravitation Group, Tata Institute of Fundamental Research, }\\\textit{Homi Bhabha Road, Mumbai - 400 005, India}\\unni@tifr.res.in}
\date{}
\maketitle

\begin{abstract}
It is pointed out that one of the systematic effects that can affect the
measurement of the speed of neutrinos significantly is the variability of the
unaveraged measurement of the distance between two points on the earth due to
thermal expansion. Possible difference between estimates done with surface GPS
apparatus and the true underground baseline can change substantially the
statistical significance of the result of superluminal speed of neutrinos,
reported recently.

\end{abstract}

While it is reasonable to believe that no particle can be accelerated through
the speed-of-light barrier to beyond the speed of light, it is not obvious
that there could not be particles that travel always with speed larger than
the speed of light, with no possibility of `decelerating' them to below the
speed of light \cite{metarelativity}. Whether there are fundamental particles
that travel at speeds beyond that of light can only be answered from careful
experiments. A recent long baseline neutrino experiment that has results
consistent with a faster-than-light neutrino \cite{neutrino} is therefore very
significant and need to be carefully examined.

The result reported in \cite{neutrino} is that the muon neutrinos seem to
travel faster than light by a small increment amounting to $(v-c)/c\simeq
2.5\times10^{-5}$ with combined errors of about $4\times10^{-6}.$ Referred to
the base line of about 730 km, this amounts to 60 ns, or a spatial distance of
18 metres. Therefore, if the accurately measured baseline has an unaccounted
systematic error of 1-2 parts in $10^{-5},$ the statistical significance of
the result can be affected drastically. Curiously, this is the level of
systematic error than can happen due to differential thermal expansion of
relevant baselines. If the distance determination was done with a bias towards
thermally expanded baseline on the surface of the earth for some reason and if
the experiment in the underground tunnel corresponds to a true average
baseline that is different from the baseline determined with GPS receivers on
surface, the anomaly reduces significantly. For example, with a conservative
thermal expansion coefficient of $7\times10^{-6}$ for the surface layers and a
difference in temperature of $3^{o}C,$ the result becomes consistent with
neutrinos travelling below the speed-of-light barrier. In any case, even a
modest temperature change of 50 millidegree C results in a change in the
baseline that is \emph{larger than the quoted error} of baseline determination
in the experiment.

It is not easy to estimate the thermal expansion due to temperature variations
over a long baseline consisting of different types of soil and rocks,
especially if there are micro-cracks in the soil. However, this can be
measured using the GPS and the variations over different time scales can be
accounted for. Inside the tunnel where the experiments are performed, as well
as along the whole baseline deep inside the earth the temperature variations
are expected to be well below $0.1^{o}C.$ So, the relevant quantity is the
\emph{difference} between baseline measured with GPS instruments that are
relatively near the surface of the earth, where temperature variations can be
large, and the estimate of the actual deep underground baseline. The
experiment was done in averaging mode that average variations over days and
also over seasons. Hence, repeated measurements on the baseline covering
various possibilities of temperature variations are also required for
eliminating doubts on possible systematic effects arising from thermal
expansion effects.

To directly detect propagation delays due to thermally generated variations of
the underground baseline, one may need statistical accuracies better than
$\ 10^{-7}$ (%
$<$%
0.3 ns over a 1000 km baseline).

\end{document}